\documentclass[a4paper,
    12pt,
    abstracton,
    ]{scrartcl}

\usepackage{amsmath} 
\usepackage{amssymb} 
\usepackage{url}
\usepackage{color}
\usepackage{hyperref}
\definecolor{refcolor}{RGB}{0,0,190}
\hypersetup{
    colorlinks,
    citecolor=refcolor,
    filecolor=refcolor,
    linkcolor=refcolor,
    urlcolor=refcolor
}
\usepackage{natbib}

\def\({\left(}
\def\){\right)}

\newcommand{\tn}{\textnormal}

\newcommand{\mc}[1]{\mathcal{#1}}

\newcommand{\abs}[1]{\left|#1\right|}

\newcommand{\schrod}{Schr\"odinger}
\newcommand{\bra}[1]{\langle#1|}
\newcommand{\ket}[1]{|#1\rangle}
\newcommand{\braket}[2]{\langle#1|#2\rangle}
\newcommand{\expectation}[1]{\langle#1\rangle}

\newcommand{\cket}[1]{||#1\rangle\rangle}

\def\sref #1{\S\ref{#1}}

\begin{document}

\author{Ovidiu Cristinel Stoica
\thanks{Department of Theoretical Physics, National Institute of Physics and Nuclear Engineering -- Horia Hulubei, Bucharest, Romania. Email: \href{mailto:cristi.stoica@theory.nipne.ro}{cristi.stoica@theory.nipne.ro},  \href{mailto:holotronix@gmail.com}{holotronix@gmail.com}}
}

\title{The universe remembers no wavefunction collapse}

\maketitle

\setcounter{tocdepth}{2}
\tableofcontents


\begin{abstract}
Two thought experiments are analyzed, revealing that the quantum state of the universe does not contain definitive evidence of the wavefunction collapse.
The first thought experiment shows that unitary quantum evolution alone can account for the outcomes of any combination of quantum experiments. This is in contradiction with the standard view on quantum measurement, which appeals to the wavefunction collapse, but it is in full agreement with the \emph{special state} proposal (L.S. Schulman in Phys Lett A 102(9):396–400, 1984; J Stat Phys 42(3):689–719 1986; Time’s arrows and quantum
measurement, Cambridge University Press, Cambridge 1997) that there are some rare states that account for quantum experiments by unitary evolution alone. The second thought experiment consists in successive measurements, and reveals that the standard quantum measurement scheme predicts violations of the conservation laws. It is shown that the standard view on quantum measurements makes some unnecessary assumptions which lead to the apparent necessity to invoke wavefunction collapse. Once these assumptions are removed, a new hope of a measurement scheme emerges, which, while still having open problems, is compatible with both the unitary evolution and the conservation laws, and suggests means to experimentally distinguish between full unitarity and discontinuous collapse.
\end{abstract}


\section{Introduction}
\label{s:intro}

As the \emph{standard quantum measurement scheme} goes \citep{vonNeumann1955foundations,Dirac58PPQM},
a measurement apparatus of an observable -- represented by a self-adjoint operator $\hat{\mc O}$ --
of a quantum system represented by $\ket{\psi}$ has a pointer state $\ket{\eta}_\lambda$ for each eigenvalue $\lambda$ of the observable $\hat{\mc O}$, and an additional state $\ket{\eta}_{\tn{ready}}$, representing the apparatus ready to perform the observation.
If the observed system is prior to the measurement in an eigenstate $\ket{\psi}_\lambda$ of $\hat{\mc O}$, then the measurement is described by the unitary process:
\begin{equation}
\label{eq:measurement-eigenstates}
\ket{\psi}_\lambda \ket{\eta}_{\tn{ready}} \stackrel{\tn{U}}{\longrightarrow} \ket{\psi}_\lambda \ket{\eta}_\lambda.
\end{equation}
The \emph{principle of superposition} leads, therefore, to the conclusion that if the observed system is in an arbitrary state $\ket{\psi}_{\tn{init}}=\sum_{\lambda}c_\lambda\ket{\psi}_\lambda$, it evolves according to
\begin{equation}
\label{eq:measurement-no-eigenstates}
\ket{\psi}_{\tn{init}} \ket{\eta}_{\tn{ready}} \stackrel{\tn{U}}{\longrightarrow} \sum_\lambda c_\lambda \ket{\psi}_\lambda \ket{\eta}_\lambda.
\end{equation}
However, since the state of the observed system is always found to be an eigenstate of $\hat{\mc O}$, one appeals to the \emph{wavefunction collapse} to project the total state to $\ket{\psi}_\lambda \ket{\eta}_\lambda$ for some outcome $\lambda$.

While this solution seems to account for all quantum experiments, the wavefunction collapse is in contradiction with the description of the evolution of the system as offered by the {\schrod} equation, which is \emph{unitary} and also has never been contradicted directly by experiments.

Von Neumann proposes a more detailed account of the quantum measurement process, by including the interaction Hamiltonian between the measurement apparatus and the observed system:
\begin{equation}
\label{eq:von-neumann-interaction-hamiltonian}
\hat{H}_{\tn{int}}(t) = g(t)\hat{\mc O}\hat{\mc P}_d.
\end{equation}
Here, the coupling $g(t)$ vanishes for all times $t$ except for the brief interval during measurement and makes the interaction Hamiltonian to dominate the process during that time. The momentum $\hat{\mc P}_d$ is the canonical conjugate of the position operator $\hat{\mc Q}_d$ of the pointer of the measurement device, so $[\hat{\mc Q}_d,\hat{\mc P}_d] = i\hbar$.
The total Hamiltonian of the observed system and the measurement apparatus is
\begin{equation}
\label{eq:von-neumann-total-hamiltonian}
\hat{H}(t) = \hat{H}_{\tn{obs. sys.}}(t) + \hat{H}_{\tn{app.}}(t) + \hat{H}_{\tn{int.}}(t).
\end{equation}
This Hamiltonian provides the unitary evolution $U$ appearing in the Eqs. \eqref{eq:measurement-eigenstates} and \eqref{eq:measurement-no-eigenstates}, and is followed by the wavefunction collapse.

In Sect. \sref{s:anamnesis}, a thought experiment is proposed and analyzed, leading to the conclusion that unitary evolution according to the {\schrod} equation can describe the results of any combinations of quantum experiments. This contradicts the point of view that the wavefunction collapse occurs with necessity. 

To find out if there is something wrong with the standard quantum measurement scheme, in Sect. \sref{s:conservation}, successive measurements are considered. It is concluded that the standard quantum measurement scheme predicts violations of the conservation laws. If the conservation laws are true, it is argued that the standard quantum measurement scheme should be revised.

In Sect. \sref{s:measurement} some implicit or explicit assumptions of the standard quantum measurement scheme are identified, showing that they are unjustified idealizations or generalizations. A truly instrumentalist formulation of quantum measurement is proposed, that is free of these assumptions, and at the same time is compatible with the conservation laws and the unitary evolution presented in the previously discussed experiments.

The problem of the wavefunction collapse and the possibility that unitary evolution alone may be enough to account for all possible quantum experiments are discussed in Sect. \sref{s:unitary}.

The arguments constructed here are perfectly compatible with and support Lawrence S. Schulman's proposal that unitary evolution is maintained at the level of a single world, based on ``special states'', which do not evolve into ``grotesque states'' like Schr\"odinger's cat \citep{schulman1984definiteMeasurements}.
The idea to resolve the measurement problem by using unitarily evolving \emph{special states} of both the observed system and the measurement apparatus was proposed in \citep{schulman1984definiteMeasurements}, and developed in \citep{schulman1986deterministicQuantumEvolution}. A review can be found in \citep{schulman1991definiteQuantumMeasurements} and an exposition in \citep{schulman1997timeArrowsAndQuantumMeasurement}.
Schulman builds models of decay, of spin boson models, cloud chamber models, and Bohm's version of the EPR experiment. The Born rule is proposed to emerge from counting the special states, using as a ``kick distribution'' the \emph{Cauchy distribution} (see \citep{schulman1997timeArrowsAndQuantumMeasurement} and references therein). Schulman calls ``kicks'' the changes of the observed state due to the interaction with the measurement device, which both should be in very special states. Despite the advances in understanding how this may happen, unfortunately, the problem was not completely settled down so far. Experiments based on modified Stern-Gerlach setups or on polarization, aiming to prove a Cauchy distributions of the kicks present in Schulman's suggested implementation of the special state approach, were proposed in \citep{schulman2012experimentalTestForSpecialState,schulman2016specialStatesDemanForceObserver,schulman2016lookingSourceChange}. As a justification for the election of the special states, Schulman proposes in \citep{schulman1989remoteTwoTimeBoundaryConditions} a final boundary condition for the universe such as Big Crunch (which at this time seems less likely to occur, because the expansion of the universe seems to be accelerated \citep{RIE98,PER99}, but who knows?), although it is not clear how final boundary conditions of the universe may explain the emergence of special pointer states in experiments.

Related to the possibility of accounting for the definite results of quantum measurements by unitary evolution alone, I proved that this always requires very special initial conditions of the observed system and the measurement apparatus \citep{Sto12QMb}, that it is possible to account not only for the definite outcomes of single measurements, but also for multiple noncommuting measurements \citep{Sto08b,Sto16aWavefunctionCollapse},
that such special states may be explained by a ``global consistency condition'' \citep{Sto12QMc,Sto13bSpringer},
and that the selection of the special states can be seen not necessarily as retrocausal, but as a continuous reduction of a set of special states with each new experiment \citep{Sto16aWavefunctionCollapse}.
These results also do not explain yet the Born rule.
To explain the emergence of pointer states, I proposed that a possible reason is a superselection rule given by the interactions involved in detection, which can be tested experimentally \citep{Sto16MicroClassCats}.

This article aims to advance in this direction by providing an argument that any combination of measurements can be described by unitary evolution alone, at the level of a single world, and that breaking unitarity by a discontinuous collapse leads to violations of conservation laws. It also proposes a generic alternative to the standard measurement scheme, but the complete details are still an open problem, due to the large number of particles involved in the measurement process.

Comparison with other approaches sharing common ideas is made in Sect. \sref{s:comparison}.
Some future developments, possible experiments, and open problems are discussed in Sect. \sref{s:future}.

\section{The unitary anamnesis experiment}
\label{s:anamnesis}

The quantum thought experiment proposed and analyzed here is rather a meta-experiment, since it can be performed on top of any combination of quantum experiments.
It aims to test the necessity of appealing to the wavefunction collapse to explain the definite outcomes of quantum experiments.

Let us assume the following setup.
\begin{enumerate}
	\item 
Various quantum experiments are performed.
	\item 
The experiments, including the preparation and the results, are recorded classically.
	\item 
The records are centralized on a server, which is a classical computer.
\end{enumerate}

Here, by \emph{classical}, I understand the level of reality that we call ``classical'', although it emerges, of course, from the quantum level, by a mechanism yet to be understood (and which will not be discussed here). Therefore, this level, sometimes called \emph{macroscopic level}, is actually still quantum, but it behaves rather similar to a classical system.
Accordingly, a measurement apparatus is a classical object, even though quantum interaction is involved during the measurement process.

The \emph{recordings} contain records of the data of the devices involved in the experiment, video recordings of the experiments, documents and reports filled by the scientists performing the experiments, anything that constitutes proof that an experiment was performed, and that certain outcome was obtained. To make the things clearer, one can imagine even that everywhere, in particular in the laboratories, there are video cameras recording everything at the classical level. All these recordings are uploaded on a central computer -- the \emph{server}.

Let us denote by $\ket{\Psi(t)}$ the state of the universe at the time $t$. The state of the universe at a time $t_{\tn{final}}$, after all quantum experiments were performed, and all the records were collected on the server, is $\ket{\Psi(t_{\tn{final}})}$. 
This state is never truly known or accessible to the observers, being themselves part of the universe, and there are in fact infinitely many such possible states, so the corresponding possible states $\ket{\Psi(t)}$ which will evolve at the time $t_{\tn{final}}$ into $\ket{\Psi(t_{\tn{final}})}$ are also infinitely many and we cannot distinguish among them objectively.

Let us apply the unitary evolution operator $U$ of the universe backward in time, to find the state of the universe $\ket{\Psi(t)}$ at any time $t\leq t_{\tn{final}}$. We get
\begin{equation}
\label{eq:backward_unitary_evolution}
\ket{\Psi(t)} = U^{-1}(t_{\tn{final}},t)\ket{\Psi(t_{\tn{final}})}.
\end{equation}
Note that $\ket{\Psi(t)}$, obtained by unitary evolution alone, is the state of the universe at the time $t\leq t_{\tn{final}}$, assuming that the evolution was unitary.
If wavefunction collapses really took place during some of the quantum experiments, then the state of the universe at the time $t$ was not necessarily $\ket{\Psi(t)}$, even though the recordings are the same in both cases.

The state $\ket{\Psi(t)}$ is a valid state, but does it make physical sense?
When a physically realistic state is evolved unitarily forward in time and a quantum measurement process takes places, the {\schrod} equation usually leads to a grotesque state -- a superposition of all possible outcome states of the measurement apparatus, a {\schrod} cat.
Therefore, should we not expect the same to happen with the state $\ket{\Psi(t)}$, obtained by the backward in time unitary evolution of the physically realistic state $\ket{\Psi(t_{\tn{final}})}$?
Most likely yes, but, on the other hand, we can check the large collection of records we have, and see no indication that $\ket{\Psi(t)}$ was grotesque.
It may be grotesque, but, since we recorded all observations and collected them, so that we can access them by inspecting (what we know about) the state $\ket{\Psi(t_{\tn{final}})}$, and there is no observable indication of $\ket{\Psi(t)}$ being grotesque, then, for all practical purposes, it was not.
Note that it is not the ambition of this article to explain why we never observe grotesque states, but only to show that (1) wavefunction collapse would lead to violations of conservation laws, and (2) that fortunately it is possible to account for the observations without making use of the wavefunction collapse.

Any state $\ket{\Psi(t)}$ obtained by applying Eq. \eqref{eq:backward_unitary_evolution} is a valid state, and the collection of states $\ket{\Psi(t)}$ accounts for all the records of the experiments by unitary evolution alone. If the server record at the time $t_{\tn{final}}$ of a particular experiment says that the pointer of the measurement apparatus reported at the time $t$ that the spin of a particle was \emph{up}, then by applying backward in time the unitary evolution operator $U$ one should obtain that, indeed, the pointer of the measurement apparatus reported \emph{up} at the time $t$.

Since the resulting outcomes of the experiments are contained among the recordings at the classical level, and also since all systems involved are quantum, applying the unitary evolution backward in time gives a unitary history consistent with all the observations.
Therefore, unless the final state $\ket{\Psi(t_{\tn{final}})}$ appeared from quantum or thermodynamical fluctuations in such a way as to merely give the illusion that these recordings represent something that really took place (similar to \emph{Boltzmann brains} \citep{schulman1997timeArrowsAndQuantumMeasurement} p. 154, \citep{albrecht2004BoltzmannBrain}), the thought experiment proposed here reveals that there is a possible unitary history of the universe, \emph{i.e.} no collapse involved, which gives the same records of the quantum experiments as the one involving collapse. This means that the appeal to wavefunction collapse is unnecessary.

This seems to be at odds with the standard quantum measurement scheme, which claims the necessity to appeal to wavefunction collapse. I will return to this in Sections \sref{s:measurement} and  \sref{s:unitary}.

This analysis also shows an apparent asymmetry between the evolution of a system forward in time, and the backward time evolution. When a quantum system is measured, the total system, containing the observed system and the measurement apparatus, seems to evolve, if unitary evolution is considered, into a superposition of classical states of the apparatus, one for each possible outcome of the measurement, as described in Eq. \eqref{eq:measurement-no-eigenstates}. Hence, the apparent necessity to appeal to collapse. By contrast, when applying the time evolution backward in time as in the proposed thought experiment, the system does not evolve into a superposition manifest at the classical level, since this would contradict the recordings collected on the server.

How is it possible? Which argument is right, the one proposed here, or the usual argument supporting the wavefunction collapse?
In Sect. \sref{s:conservation} I will try to elucidate this problem.

\section{The conservation laws experiment}
\label{s:conservation}

The second thought experiment, discussed in this section, refers to an apparently different problem -- that of the \emph{conservation laws}. We will see that the standard quantum measurement scheme is in contradiction with the conservation laws.

In quantum mechanics, an observable quantity is \emph{conserved} if the corresponding self-adjoint operator $\hat{\mc O}$ commutes with the Hamiltonian $\hat{\mc H}$ from the {\schrod} equation.
However, the wavefunction collapse is described by projection operators, which do not commute simultaneously with all the conserved quantities.
This means that if the wavefunction collapse really takes place, violations of conservation laws should appear.
Here, a spin is put on the violation of the angular momentum conservation due to the wavefunction collapse.

The following thought experiment aims to test this prediction for spin measurements and conservation of the angular momentum.
It is based on a thought experiment present in the literature, and was also used by Wigner to show that some measurements cannot be sharp because of the conservation laws \citep{wigner1952MessungQMOperatoren,busch2010EnTranslationWigner1952MessungQMOperatoren}. Similar conclusions can be obtained from \citep{schulman2016specialStatesDemanForceObserver}, where spin measurements are discussed as well.
Another proof of the tension between wavefunction collapse and conservation laws is given in \citep{Burgos1993ConservationLawsQM}.

Consider two noncommuting observables $\hat{\mc O}_1$ and $\hat{\mc O}_2$ that commute with the Hamiltonian, hence, they are conserved.
For example, they can be the spin components of a $\frac 1 2$-spin particle, in two different directions $x$ and $z$. This is a variation of a standard thought experiment involving spin measurements, with an emphasis added on the angular momentum conservation.
For practical reasons, it is better that the observed particle is electrically neutral, but has a nonzero magnetic moment, for example a neutron or a silver atom, but in principle one can also consider electrons.

We first measure the spin along the $x$ axis, resulting in one of the possible eigenstates $\ket{\uparrow}_x$ and $\ket{\downarrow}_x$. Let us assume that the outcome is $\ket{\uparrow}_x$. Then, we measure the spin along the $z$ axis, resulting with equal probability in one of the states $\ket{\uparrow}_z$ and $\ket{\downarrow}_z$. If we measure it again along the $x$ axis, there is a $50\%$ chance to obtain the result $\ket{\downarrow}_x$. Obviously, for the observed particle the spin was not conserved. The only way out, if we find unacceptable that angular momentum conservation is broken, is to assume that the total angular momentum of the overall system, including the spin of the observed particle and the angular momentum of the measurement devices, is conserved.

However, the standard measurement scheme does not allow us to assume a transfer of angular momentum between the observed system and the measurement devices such that the total angular momentum is conserved!

Consider the Stern-Gerlach devices constructed in such a way that they have three possible states: $\ket{\tn{ready}}$, $\ket{\tn{up}}$, and $\ket{\tn{down}}$, and that the state $\ket{\tn{ready}}$ has a certain angular momentum, say $0$. While a Stern-Gerlach device contains magnets, made of atoms with magnetic moment, like iron, it is possible to conceive such devices with the total angular momentum $0$. For example one can add to them some parts that have the only role of canceling the angular momentum.

Suppose the Stern-Gerlach device is prepared to measure a particle's spin along the $x$ axis, and that the particle's spin is already an eigenstate of the spin operator $\hat{\mc S}_x$ along the $x$ axis. Then the measurement does not change the spin of the observed particle, and the total angular momentum is conserved. The process is
\begin{equation}
\ket{\uparrow}_x \ket{\tn{ready}}_x \stackrel{\tn{U}}{\longrightarrow} \ket{\uparrow}_x \ket{\tn{up}}_x.
\end{equation}
The angular momentum conservation implies that the total angular momentum of $\ket{\tn{up}}_x$ should be the same as the total angular momentum of $\ket{\tn{ready}}_x$, and as mentioned we arranged for it to be $0$. Likewise for the total angular momentum of $\ket{\tn{down}}_x$.

But if one measures the spin along the $z$ axis, the standard measurement scheme predicts that by unitary evolution, the system evolves like this
\begin{align}
\ket{\uparrow}_x \ket{\tn{ready}} &= \frac{1}{\sqrt{2}}\(\ket{\uparrow}_z + \ket{\downarrow}_z\) \ket{\tn{ready}}_z \\
&\stackrel{\tn{U}}{\longrightarrow} \frac{1}{\sqrt{2}}\(\ket{\uparrow}_z  \ket{\tn{up}}_z + \ket{\downarrow}_z \ket{\tn{down}}_z\),
\end{align}
and the final state is obtained by projection, so it is either $\ket{\uparrow}_z \ket{\tn{up}}_z$ or $\ket{\downarrow}_z \ket{\tn{down}}_z$, with equal chances.

Hence, according to the standard quantum measurement scheme, there is no trace of the initial value of the spin. This means that, by combining the successive measurements along the axes $x$, $z$, and $x$ again, one may obtain a violation of the angular momentum conservation. By indexing the three Stern-Gerlach devices with $1,2,3$, and assuming the initial spin state of the particle to be $\ket{\uparrow}_x$, and the final pointer states respectively $\ket{\tn{up}}_{x1},\ket{\tn{up}}_{z2},\ket{\tn{down}}_{x3}$ (which happens in $25\%$ of the cases), the total system evolves as follows. During the first measurement it evolves like
\begin{align}
\ket{\uparrow}_x \ket{\tn{ready}}_{x1}\ket{\tn{ready}}_{z2}\ket{\tn{ready}}_{x3}\\
\stackrel{\tn{U}}{\longrightarrow} 
\ket{\uparrow}_x \ket{\tn{up}}_{x1}\ket{\tn{ready}}_{z2}\ket{\tn{ready}}_{x3}.
\end{align}
During the second measurement, the evolution of the total system is
\begin{align}
\ket{\uparrow}_x \ket{\tn{up}}_{x1}\ket{\tn{ready}}_{z2}\ket{\tn{ready}}_{x3}\\
\stackrel{\tn{U,Pr}}{\longrightarrow} 
\ket{\uparrow}_z \ket{\tn{up}}_{x1}\ket{\tn{up}}_{z2}\ket{\tn{ready}}_{x3}.
\end{align}
During the third measurement, the evolution is
\begin{align}
\ket{\uparrow}_z \ket{\tn{up}}_{x1}\ket{\tn{up}}_{z2}\ket{\tn{ready}}_{x3}\\
\stackrel{\tn{U,Pr}}{\longrightarrow} 
\ket{\downarrow}_x \ket{\tn{up}}_{x1}\ket{\tn{up}}_{z2}\ket{\tn{down}}_{x3}.
\end{align}

Hence, since the angular momentum is $0$ for each of the three devices in each of their states, the standard measurement scheme predicts an angular momentum difference of $1$ between the initial and final total states.

In order for the angular momentum to be conserved for the total system, the three measurement devices should contain the missing angular momentum along the $x$ axis. In fact, any measurement device should be able to preserve the difference of any conserved quantity $\hat{\mc A}$ of the observed system which is changed during the measurement process.

Remember that for an observable $\hat{\mc A}$ commuting with the Hamiltonian, not only the eigenstates of the observable are conserved, but also the quantities of the form $\bra{\psi}\hat{\mc A}\ket{\psi}$, called the \emph{expectation value} of $\hat{\mc A}$ in the state $\ket{\psi}$ (although strictly speaking the term ``expectation value'' should refer only to the probabilities, in conjunction with the \emph{Born rule}). This means that the measurement apparatus should contain in its quantum state not only the difference of the observable it measures, but also of the expectation values of any conserved quantity $\hat{\mc A}$ in the state $\ket{\psi}$.

But if the measurement devices contain the missing expectation values of the angular momentum along all possible directions, the final state actually contains the complete information about the initial state $\ket{\psi}_{\tn{init}}$. Suppose the expectation values of the spin along the three axes, in a state $\ket{\psi}$, are $s_x=\expectation{\hat{\mc{S}}_x}_{\ket{\psi}}$, $s_y=\expectation{\hat{\mc{S}}_y}_{\ket{\psi}}$, and $s_z=\expectation{\hat{\mc{S}}_z}_{\ket{\psi}}$. Then in the \emph{Bloch sphere representation} the spin state $\ket{\psi}$ is represented as a unit vector in the Euclidean space, having as Cartesian components just $(s_x,s_y,s_z)$. Hence, in the experiment analyzed in this section, angular momentum conservation implies that the initial state is never lost. Spin measurements are completely reversible, while projections should completely destroy the information about the initial state.

This analysis invites us to reconsider the standard measurement scheme, and to search for a better description of quantum measurements.

\section{What we really know about quantum measurement?}
\label{s:measurement}

We do not know yet what quantum interactions constitute a quantum measurement, and this article does not claim to provide the answer.
However, the two thought experiments discussed in the previous sections suggest that the standard quantum measurement scheme needs to be revised.
Bohr's own view and the Copenhagen Interpretation divide the world into the observed system, which is quantum, and the measurement device, which is seen as classical for all practical purposes, and macroscopic. 
What does it mean for a quantum system to be classical is still an open problem, so it is taken as it is given -- just a measuring device able to distinguish between eigenvalues of a property of a quantum system.
In the absence of such a characterization, in the following I will also just take it as given, and only assume that (1) a macroscopic object is still quantum,
and (2) that by knowing how it appears to us macroscopically cannot tell us how it is at a detailed quantum level, but it can constrain what we know up to an equivalence class. This is akin to the approach to thermodynamics based on coarse graining of the phase space, where the laws governing the detailed dynamics is assumed to be known, but not the detailed state.
Having a precise definition of a measurement device and of a quantum state that appears classical at the macroscopic level would be desirable, but it is lacking. However, we can assume for the moment that a quasi-classical system is just a collection of quantum states that are indistinguishable at the macroscopic level, leaving it for future research to find the precise definition.

The discussion in Sect. \sref{s:conservation} led us to the conclusion that the conservation laws would be violated, unless the final state of the apparatus contains somehow the missing parts of all conserved quantities of the observed system. In the example of angular momentum and spin, one would expect that the final state of the apparatus contains the difference of angular momentum.
If indeed the missing quantity is contained in the final state of the apparatus, it still should not be visible at the classical level, where the way the apparatus looks depends only on the outcome $\lambda$.
While the various final states of the apparatus should appear classically to depend only on $\lambda$, the quantum state has to also depend on the initial state of the observed system. This is indeed possible, because the apparatus is a macroscopic quantum system, and macroscopically one cannot distinguish among its various possible quantum microstates.
One can see this as similar to the \emph{decoherence approach} to quantum measurements, if we regard the internal degrees of freedom of the apparatus as the environment (see Sect. \sref{s:comparison}).

Let us denote the final state of the apparatus as a function of both the initial state $\ket{\psi}_{\tn{init}}$ and the final state $\ket{\psi}_\lambda$ of the observed quantity,
\begin{equation}
\ket{\eta}_{\lambda,\ket{\psi}_{\tn{init}}}.
\end{equation}
If the observed system was initially in the state $\ket{\psi}_{\tn{init}}=\ket{\psi}_\lambda$, then the final state of the apparatus is $\ket{\eta}_{\lambda,\ket{\psi}_\lambda}$.
However, in general, the final state of the measurement apparatus is microscopically different for different initial states $\ket{\psi}$.

We introduce a relation of equivalence $\cong$, called \emph{macroscopic equivalence}. Two states of a quantum system are macroscopically equivalent if they cannot be distinguished by classical observations. Applied to our measurement apparatus, we have
\begin{equation}
\ket{\eta}_{\lambda,\ket{\psi}_{\tn{init}}} \cong \ket{\eta}_{\lambda,\ket{\psi}_\lambda}
\end{equation}
for any initial state $\ket{\psi}_{\tn{init}}$ of the observed system, as long as the final state is $\ket{\psi}_\lambda$. This expresses the fact that macroscopically we can only know what the apparatus says about the observable $\hat{\mc O}$ of the observed system, hence its final state $\ket{\psi}_\lambda$, and not its real state $\ket{\psi}_{\tn{init}}$ prior to the measurement.
In addition, for any two possible initial states of the observed system, $\ket{\psi_1}$ and $\ket{\psi_2}$,
\begin{equation}
\ket{\eta}_{\lambda_1,\ket{\psi_1}} \cong \ket{\eta}_{\lambda_2,\ket{\psi_2}}
\end{equation}
if and only if $\lambda_1=\lambda_2$.

Since many different quantum states may look similar macroscopically, one can define a \emph{macroscopic state} as being the \emph{equivalence class} of the equivalence relation $\cong$. We will denote the macroscopic equivalence class of a quantum state $\ket{\eta}$ by $\cket{\eta}$.

Let us describe again what a quantum measurement is, this time emphasizing that the measurement device is classical, and its states are macroscopic.
Now the state of the system is not considered to be completely known at the quantum level. Since the relation of macroscopic equivalence is not preserved by the unitary evolution, the results have a degree of unpredictability.

The equation \eqref{eq:measurement-eigenstates} should be replaced by the following:
\begin{equation}
\label{eq:measurement-eigenstates-improved}
\ket{\psi}_\lambda \cket{\eta}_{\tn{ready}} \longrightarrow \ket{\psi}_\lambda \cket{\eta}_\lambda,
\end{equation}
which apparently is quite predictable, although the complete quantum state of the measurement apparatus is not known.

If the observed system is in an arbitrary state $\ket{\psi}_{\tn{init}}=\sum_{\lambda}c_\lambda\ket{\psi}_\lambda$, then the evolution is unpredictable both at quantum and at classical level, since the unitary evolution does not preserve the macroscopic equivalence.
The experiments tell us that
\begin{equation}
\label{eq:measurement-no-eigenstates-improved}
\ket{\psi}_{\tn{init}} \cket{\eta}_{\tn{ready}} \longrightarrow \ket{\psi}_\lambda \cket{\eta}_\lambda,
\end{equation}
where any eigenvalue $\lambda$ can be obtained, with the probability $\abs{c_\lambda}^2$.

We do not know why the final macrostate of the apparatus has to be in one of the states $\cket{\eta}_\lambda$, and not in a superposition as in Eq. \eqref{eq:measurement-no-eigenstates}. In fact, superpositions also do not commute with the macroscopic equivalence, so one cannot know how a superposition of macroscopical states looks like, based on the macroscopical states alone.
What we know is that the measurement apparatus is always in one of the states $\cket{\eta}_\lambda$ at the end of a measurement.
In addition, we know that the probabilities are $\abs{c_\lambda}^2$ according to the Born rule, but we do not have an explanation for this.

Equation \eqref{eq:measurement-no-eigenstates-improved} should be used instead of \eqref{eq:measurement-no-eigenstates}  to describe the measurement process, because it captures our lack of knowledge of the quantum state of the measurement apparatus. It also explains why it is possible to have unitary descriptions of the measurement process, without having to appeal to the wavefunction collapse. However, it does not imply that there is no wavefunction collapse, it just shows that the collapse does not necessarily happen.

The standard measurement scheme tacitly identifies the macroscopic equivalence of the states of the measurement apparatus with their quantum equivalence.
In addition, as we have seen, during quantum measurement all the conserved quantities should be exchanged between the observed system and the measurement apparatus, to be conserved. This strongly invites us to revise the standard quantum measurement scheme.
To save the idea of the wavefunction collapse, one would have to propose a mechanism by which the collapse of the observed system is correlated with that of the apparatus, such that the conservation laws are still valid.
Given that the reason these observables are conserved is their commutativity with the Hamiltonian, it is much simpler to admit that, even during measurements, the unitary evolution should remain valid. The apparent collapse should in fact be understood as a unitary quantum process which includes interactions between the observed system and the measurement apparatus.


The instrumentalist point of view on quantum mechanics, usually referred to as the \emph{orthodox interpretation} of quantum mechanics, states that one should only refer to observable quantities, which are what the measurement apparatus displays.
Any additional assumptions should be avoided.

Precisely by taking the instrumentalist point of view seriously here was revealed that the standard measurement scheme is based on hidden assumptions. Once these assumptions removed, we see that there is no evidence supporting the wavefunction collapse, and the unitary evolution is compatible with all our experiments, as Sect. \sref{s:anamnesis} shows.

The main assumption identified was the one that the quantum state of the measurement apparatus depends only on the outcome $\lambda$. Another assumption, discussed in \citep{Sto16aWavefunctionCollapse}, is that if the apparatus reports that the outcome is $\lambda$, then the state of the observed system is an eigenstate corresponding to the eigenvalue $\lambda$. This is not true in the case of position measurements for photons, since any such measurement results in the photon being absorbed by an electron of an atom. However, as the {\schrod} equation for the atom shows, the wavefunction of the electron extends to infinity, despite having a bump in the proximity of the nucleus, so the position of the absorbed photon is only approximately localized. This means that in practice one cannot measure the position with a complete accuracy. The same goes for other measurements, which consist in exciting atoms. In addition, spin measurements rely on interactions of the observed particle with the magnetic field:
\begin{equation}
\label{eq:spin-interaction-hamiltonian}
\hat{H}_{\tn{int}}(t) = -\mu\mathbf{S}\cdot\mathbf{B},
\end{equation}
where $\mu\mathbf{S}$ is the magnetic moment of the observed particle, and $\mathbf{B}$ is the magnetic field.
The spin of the observed particle will be disturbed, unless it was initially aligned along the magnetic field.
In the standard quantum measurement scheme, this Hamiltonian can be plugged into Eq. \eqref{eq:von-neumann-interaction-hamiltonian}, and then the collapse should be applied. However, in order for the angular momentum to be conserved, one should also properly take into account the backreaction of the observed particle on the measurement apparatus. This should be completed with an explanation of why the final state of the apparatus is always a pointer state. We do not have yet a proper mechanism for this.

Taking a purely instrumentalist position does not amount to claiming that everything is properly described by the orthodox interpretation and the standard quantum measurement scheme. By contrary, it requires eliminating any assumptions that were not derived directly from experiment or that cannot be tested by experiments, directly or through their predictions. And once removed such assumptions, one remains with the modest conclusion that we do not actually know what a measurement is, why the macroscopic world appears to behave classically, and whether the wavefunction collapse exists or not.

To truly be instrumentalists, one should formulate the quantum measurement scheme by making as few unnecessary assumptions as possible. The following formulation aspires to be a step forward in this direction:

\begin{enumerate}
	\item 
A measurement apparatus for an observable $\hat{\mc O}$ of a quantum system is a device having a macrostate $\cket{\eta}_{\tn{ready}}$, and a state $\cket{\eta}_{\lambda}$ at least for a subset $\Lambda$ of the eigenvalues of the observable $\hat{\mc O}$. Corresponding to the eigenvalues $\lambda\notin\Lambda$, there is a final macrostate $\cket{\eta}_{\tn{failed}}$. This is because the apparatus is not perfect, and it cannot always cover all the eigenvalues of $\hat{\mc O}$.
An example is the position measurement.
	\item 
If the final state of the apparatus is $\cket{\eta}_{\lambda}$, this means that the state of the observed system was at some time during the measurement \emph{approximately} an eigenstate corresponding to the eigenvalue $\lambda$. The apparatus should also come with a specification of the tolerances of detection, which define quantitatively what should be understood by ``approximately''.
	\item 
The observed system is not guaranteed to remain in the detected state, because on the one hand the interaction with the apparatus changes both of their states, sometimes even after detection, and on the other hand, if the observable does not commute with the Hamiltonian, the observed system will be disturbed.
	\item 
If the observed system is believed to be in a state $\ket{\psi_1}$, by previous measurements combined with the subsequent unitary evolution according to the {\schrod} equation, the probability to be reported by the measurement apparatus to be found in a state $\ket{\psi_2}$ by a subsequent measurement is given by the Born rule, $\abs{\braket{\psi_1}{\psi_2}}^2$, again taking into consideration the measurement errors and the possible disturbances.
\end{enumerate}

This measurement scheme definitely can be improved. However, at least in the present form, it leaves plenty of freedom for unitary evolution, and the appeal to the wavefunction collapse is not necessary.

\section{Does the wavefunction really collapse?}
\label{s:unitary}

Unlike the standard quantum measurement scheme, the measurement scheme proposed in Sect. \sref{s:measurement} leaves enough freedom to the measurement process to allow for the conservation laws to remain unbroken. It also does not contradict -- at least not explicitly -- the unitary evolution, although it does not prove it either.

Since the commutation of observables with the Hamiltonian is what makes them be conserved, it is possible to restore the conservation laws by assuming that unitary evolution remains valid during measurements. However, one may conceive an improved version of the standard measurement scheme, appealing to the wavefunction collapse in a smarter way, which allows for the conservation laws to hold. Admittedly, this may be very complicated, but it may be possible. 

Such an improved version of the standard measurement scheme may be along the following lines:

\begin{quote}
The observed system indeed collapses, but in the same time some particles from the measurement apparatus also collapse, in a way that ensures the validity of all conservation laws. The collapse of the particles from the apparatus happens in a way that is not observable at the classical level.
\end{quote}

This solution seems to be at least as good as the possibility of a solution based on unitary evolution alone, which is also not observable at the classical level.
However, in addition, it has some problems.

First, if we take into account special relativity, any solution should be invariant at changes of the reference frame. That is, the conservation laws should not depend on the way spacetime is sliced into spacelike subspaces of simultaneity.
In any proposed solution to make the wavefunction collapse ensure the conservation laws, one should take care that the conservation laws also hold for any such slicing of spacetime.
The only way to do this seems to be to ensure the conservation laws through local interactions between the measurement apparatus and the observed system.
A multiple collapse that is synchronized to ensure the conservation laws will always require a preferred reference frame.
Moreover, the amplitudes of the collapsing system do not sum up to $1$ in other reference frames.
By contrast, the {\schrod} equation is a \emph{partial differential equation}, which satisfies the condition that all the interactions are local.
Therefore, unitary evolution seems to be the only way to ensure the validity of conservation laws.

The second problem of an explanation that is still based on collapse is that it is not needed, once we realize from the thought experiment proposed in Sect. \sref{s:anamnesis} that unitary evolution can also explain the outcomes of any combinations of quantum measurements.

This does not mean that the unitary evolution explanation of quantum measurements is proven. First of all, its main disadvantage is that we do not know how to use it to describe the measurement process. A solution based on unitary evolution alone was proposed in \citep{Sto08b,Sto08f,Sto12QMc,Sto13bSpringer,Sto12QMb} and detailed in \citep{Sto16aWavefunctionCollapse} (where in particular nonlocality and contextuality are discussed), but the cited works only show the possibility of such a solution and analyze the potential obstacles, and not how it works effectively during quantum measurements.
Another problem is that we do not know yet how to derive the Born rule from such a unitary evolution explanation. But the upside is that at least it may be possible to allow the derivation of the Born rule, while the standard quantum measurement scheme merely postulates it in conjunction with the rather violent wavefunction collapse.
However, it is not the purpose of this article to develop a unitary evolution solution to the measurement problem. The purpose is more modest, to point some problems with the standard quantum measurement scheme, and to show that it does not preclude a unitary evolution solution.

While the measurement scheme proposed in Sect. \sref{s:measurement} mentions the Born rule, this does not mean that it assumes a wavefunction collapse. This collapse is purely \emph{epistemic}, since we do not really know the precise quantum states and their precise evolution. We only know what the measurement apparatus reports, but the evolution itself can very well take place unitarily at all times, even during the measurement. It is therefore justified to allow for the possibility that there is an \emph{ontic wavefunction} that evolves unitarily. In this case, our knowledge about this state is merely an approximation, which constitutes an \emph{epistemic wavefunction}, described so well by the standard quantum measurement scheme, and which is allowed to collapse \citep{Sto16aWavefunctionCollapse}, because the epistemic collapse is just an update of our knowledge about the ontic wavefunction. Such a position is perfectly compatible with the thought experiments analyzed in Sections \sref{s:anamnesis} and \sref{s:conservation}. The recordings stored on the server from the unitary anamnesis experiment (Sect. \sref{s:anamnesis}) are merely the kind of data from the quantum measurement scheme proposed in Sect. \sref{s:measurement}, and the unitary evolution can account for all these records.

\section{Comparison with other approaches}
\label{s:comparison}

The results presented in Sect. \sref{s:conservation} contradict the idea of a discontinuous collapse. From Sect. \sref{s:anamnesis} follows that it is possible to account for quantum measurements by unitary evolution alone, as Lawrence S. Schulman proposed in his ``special state'' approach discussed in the Introduction \citep{schulman1984definiteMeasurements,schulman1997timeArrowsAndQuantumMeasurement}.
In \citep{Sto08b,Sto08f} I proposed that by taking into account the internal degrees of freedom of the measurement devices and their quantum interactions of the observed system, it is possible for a unitary evolution of a quantum state vector to satisfy even the conditions imposed by successive measurements. I continued to develop these ideas in \citep{Sto12QMb,Sto12QMc,Sto13bSpringer,Sto16aWavefunctionCollapse}, and in the present article.

Schulman brings good arguments that the kicks should obey a Cauchy distribution, and he proposes ways to find experimental signatures of this particular implementation, which, if found, would support the special state approach \citep{schulman2012experimentalTestForSpecialState,schulman2016specialStatesDemanForceObserver,schulman2016lookingSourceChange}.
The argument presented in Sect. \sref{s:conservation} based on conserved quantities leads to a different kind of experimental evidence, which apparently we already have: if the wavefunction collapses in a discontinuous way, we should already see situations where the angular momentum and other conserved quantities are actually not conserved, yet such violations were not observed so far. However, the experiment performed so far did not actually look for violations of conservations, but now that we know what to search for, it may be possible to make new experiments targeted towards finding such violations, opening by this new possibilities for testing the proposal.

Another difference is in the proposed ways to explain the origin of the special states. Schulman relies on final boundary conditions of the universe, particularly on a \emph{big crunch} which seems to be less likely to happen after the discovery that the expansion of the universe appears to be accelerated. This does not rule out completely the possibility of final boundary conditions, because it is possible to impose such conditions on the future conformal boundary. In particular, such conditions will just be initial conditions for the future cycle of the universe in Penrose's \emph{conformal cyclic cosmology} \citep{Pen07,Pen11}. According to conformal cyclic cosmology, our universe is part of an infinite cycle of universes, each of them expanding forever, but such that its final conformal boundary state is related to the initial state of the next one by a conformal transformation.
However, the special states provided by a final boundary condition may be completely different from the special states needed to account for the pointer states obtained in quantum measurements, and it seems difficult to see how the former may lead to the latter.
Unitarity demands that both the observed system and the apparatus are in special states and ``conspire'' to give the desired result, so this is why I think that a generic final boundary condition is not likely to explain all these ``conspiracies'', which depend on the choice of the measured observables.
My position about this is to consider ``delayed initial conditions'' which are imposed during the measurements themselves \citep{Sto08f,Sto08b}. In fact, even the wavefunction collapse is of this kind, by resetting the wavefunction at various times, and not at the initial or final times, but if the evolution is truly unitary, the delayed initial conditions affect both the future and the past. These delayed initial conditions should be mutually consistent, and thus satisfy a \emph{global consistency principle}, which can be understood in terms of \emph{sheaf theory} \citep{mac1992sheaves,bredon1997sheaf} applied on the block world of special and general relativity \citep{Sto12QMc,Sto13bSpringer}.
A way to naturally impose delayed initial conditions during measurements can come in the form of a new superselection rule, which would allow superpositions of interactions resulting in decay or absorption only in certain situations. In \citep{Sto16MicroClassCats} I propose such a superselection rule associated with the interactions that usually take place during detection of particles, and several experiments aiming to test this superselection rule.

Although the central point of this article is to work within the good old standard quantum mechanics to explain away the wavefunction collapse, in the same spirit of Schulman's works, common points and differences can be established with several other alternative approaches to quantum mechanics. The literature on this subject is rich, but I will try to make justice at least to the most representative approaches.

Probably the first approach that takes unitary evolution seriously is the \emph{pilot-wave theory} (PWT) \citep{deBroglie1926OndesEtMouvements,Bohm52}, in which the pilot-wave always evolves unitarily, without collapse. Bohm made use of the decoherence mechanism to explain the apparent collapse, and of the \emph{Bohmian trajectories} to select the branch of the wavefunction corresponding to the outcome of each measurement. The \emph{many-world interpretation} (MWI) also uses decoherent branches of a single wavefunction evolving unitarily, but considers all of them equally valid as worlds \citep{Eve57,Eve73,dWEG73}.
It is true that MWI is unitary, if we take into account all single worlds in which the full wavefunction is decomposed after each measurement. Consequently, the conservation laws hold at the level of all worlds. However, at the level of each world, the collapse is still present, so the problem with the conservation laws remains for the observers embedded in each particular branch (see Sect. \sref{s:conservation}). These approaches contrast with the proposal to consider unitary evolution at the level of a single world, and without appealing to hidden variables. However, since in the unitary evolution proposal, as explained in \citep{Sto08b} (the first archived version is more detailed) and \citep{Sto16aWavefunctionCollapse}, every new constraint due to observations reduces the set of possible unitary evolving solutions of {\schrod}'s equation, we can interpret it as a kind of many-world scenario with unitarily evolving histories. Only the macroscopic description is actually branching, while at quantum level the splitting is done for the entire future and past history.

There is an important similarity between the present proposal and the \emph{decoherence approach} \citep{Zur03a,schlosshauer2006experimental}. This resides in the appeal to additional degrees of freedom, which are absent in the standard quantum measurement scheme. In the decoherence approach, these degrees of freedom are represented as the quantum state of the environment, which can be seen as including those of the measurement apparatus which are additional to the pointer state. The decoherence approach is sometimes understood as providing a unitary evolution of the quantum state, but again this is true only when all branches are accounted for. Even if decoherence due to the environment diagonalizes the density matrix of the observed system in the preferred pointer basis, the next step, which consists in interpreting the density matrix as a statistical ensemble of eigenstates of the observable, contains collapses. The reason is that two successive noncommuting quantum measurements lead to different ways to interpret the density matrix as a statistical ensemble, which cannot be accommodated by the same quantum state under unitary evolution without appealing to special states.

In \citep{moldoveanu2013unitary} the Grothendick group construction for the tensor product commutative monoid was used, together with envariance and quantum Darwinism. Notable for its position on unitary evolution is the \emph{cellular automaton interpretation} \citep{tH07,tHooft2014CellularAutomatonInterpretationQM,Elze2014Action4CellularAutomata}.

If we assume that the total quantum state evolving unitarily satisfies consistently the constraints imposed by all quantum measurements, then the evolution is deterministic. Hence, the total state has to be so that when the quantum measurements are made, definite outcomes are obtained, resulting in an apparent retrocausality. One sort or another of apparently retrocausal behavior seems to be unavoidable in any realistic model, as the Kochen-Specker theorem shows \citep{Bel66,ks65}, and as follows from Wheeler's \emph{delayed choice experiment} \citep{Whe78,delayed2007}. This is also acknowledged in the \emph{transactional interpretation} \citep{cramer1986transactional,cramer1988overview} and other approaches to quantum mechanics \citep{deBeauregard1953-DEBMQ,Rietdijk1978retroactiveInfluence,Wharton2007TimeSymmetricQM,price2008toyRetrocausality,price2015disentangling}.

The retrocausal position is fundamental in particular in the \emph{two-state vector formalism} (TSVF) \citep{aharonov1964time,aharonov1991complete}, where it is considered that the initial state vector of a quantum system contains only an incomplete description of reality (as in PWT). According to TSVF, to make the description complete, between any pair of successive measurements of the same system at times $t_{k}<t_{k+1}$, $k\geq 0$ integer, one should take into account the state vector $\ket{\psi_{k}}$, fixed by the measurement $k$ and evolving forward in time by $U(t,t_{k})\ket{\psi_{k}}$, and another state vector $\bra{\psi_{k+1}}$, fixed by the measurement $k+1$ (\emph{postselected}) and evolving backward in time by $\bra{\psi_{k+1}}U(t_{k+1},t)$.
Apparently between $t_{k}$ and $t_{k+1}$ only the state vectors $\ket{\psi_{k}}$ and $\bra{\psi_{k+1}}$ are relevant. 
This approach led to various interesting predictions of what happens in between the measurements, tested by weak value measurements, as for example in \citep{aharonov2012future-past}.

While proposals like PWT and TSVF assume the wavefunction to be incomplete and try to complete it with additional objects or variables, the usual wavefunction may be enough, but the constraints imposed if we assume that the standard quantum measurement scheme overdetermines it. The unknown quantum degrees of freedom of large objects like the measurement device or the environment complement the observed system with the necessary information to give the complete description, including the outcomes of measurements. If one wishes, these degrees of freedom can be considered similar to hidden variables, not being themselves completely accessible to observations, but they are not an addition to the degrees of freedom of the Schr\"odinger's equation. 

There are proposals to remove the apparent wavefunction collapse by using both initial and final boundary conditions, for example in \citep{sutherland2008causallySymmetricBohm} in the case of PWT, and in \citep{Sto08b} for standard quantum mechanics.
In addition, the TSVF was recently extended to include the measurement device and to provide a description of reality based on unitary evolution of a \emph{two-state density matrix}, explaining away the apparent wavefunction collapse \citep{AharonovCohen2014MeasurementCollapseTSVF,cohen2016quantum2classical}. This time the initial and final states include not only the observed system, but the measurement device as well as the environment. The initial state evolves to become a superposition of different possible states of the total system at the end of the measurement, while the backward evolving state corresponds to one of the possible resulting states of the complete system. Using a decoherence mechanism applied to the two-state density matrix, a two-state unitary model of measurement is constructed \citep{AharonovCohen2014MeasurementCollapseTSVF}. The same mechanism is proposed to explain the quantum-to-classical transition \citep{cohen2016quantum2classical}, and to resolve some problems of the many-world interpretation. 

A major difference between TSVF and the single unitary evolution approach is the employment of a single wavefunction evolving unitarily, similar to standard quantum mechanics.
The single quantum state is subject of constraints that are spread in space and time at all the places and times when measurements are performed -- the \emph{delayed initial conditions} \citep{Sto08b}, all of them satisfying a \emph{global consistency condition} to admit a unitarily evolving solution \citep{Sto12QMc}.
Each of these constraints is incomplete, and applies both forward and backward in time, because unitary evolution is deterministic.
These are logically unavoidable consequences of attempting to make the unitary evolution from standard quantum mechanics remain valid also during the apparent wavefunction collapse.

In the quantum anamnesis experiment from Sect. \sref{s:anamnesis}, because we start from the state at a time $t_{\tn{final}}$, which is evolved backward in time unitarily, one can think that there is a similarity to the state vector evolving backward in time in TSVF and other proposals of using final boundary conditions, including Schulman's. This is not the case, since the state considered in my proposal is the quantum state evolving as usual toward the future, it is only reconstructed \emph{a posteriori} from the state of the universe at $t_{\tn{final}}$, which contains the server storing all the classical information about the quantum experiments. In fact, the data stored on the server represent nothing else than the conditions imposed by the quantum measurements and their outcomes at various places and times, as well as the classical observations, which together are the recordings, the delayed initial conditions mentioned above. They are all stored and evolved forward until the time $t_{\tn{final}}$, with the single purpose to gather at the same time all the delayed initial conditions, to have a single-time condition. This condition at $t_{\tn{final}}$ plays the role of an ``initial condition'' (in the sense from the theory of partial differential equations, where there is no distinction between ``initial'' and ``final'' conditions, and conditions at any other given time), which allows one to find a collapse-free solution of the {\schrod} equation for the universe. A difference between approaches using an interplay between initial and final conditions and the approach advocated here is the following. While in \citep{sutherland2008causallySymmetricBohm} and \citep{AharonovCohen2014MeasurementCollapseTSVF} initial and final conditions are imposed on distinct state vectors to explain away the apparent collapse, and while in \citep{Wharton2007TimeSymmetricQM,wharton2010timeSymmetricBCandQM} initial and final conditions are used in a Lagrangian formulation, and while in Schulman's approach, the final boundary conditions are supposed to determine the special states, in my proposed thought experiment, the interplay is between the constraints that the classical recordings of the outcomes of measurements at various places and times impose on the unitarily evolving quantum system. This should be clear from Sect. \sref{s:measurement} and in fact from the rest of the article. A description of how this interplay of constraints should work was made in \citep{Sto12QMc,Sto16aWavefunctionCollapse}.

\section{Future research}
\label{s:future}

The arguments made here and elsewhere for an explanation of the apparent collapse by unitary evolution alone are just a beginning, and a modest addition to Schulman's approach.
More work remains to be done. One major open problem is finding a concrete mechanism implementing the quantum measurement scheme described in \sref{s:measurement}. Such a mechanism was suggested in \citep{Sto08b}. This should be complemented with an explanation of the suppression of superposition which makes measurements to have definite outcomes. Possibly, this will be solved when an explanation for the quantum-to-classical transition will be found. Another major problem is deriving the Born rule, if possible without additional principles, making use of a homogeneous probability distribution on the set of possible initial conditions, updated by the future constraints due to measurements. In the meantime, it would be helpful to try to identify departures from the predictions of standard quantum mechanics, and test them by experiments. A starting point may be to test the conservation laws, according to Sect. \sref{s:conservation}. However, it is difficult to see how we can do this for large system like the measurement apparatus, when the violations of conservation laws are so small. Finding out that the conservation laws are, indeed violated as predicted by the standard quantum measurement scheme would be the ultimate evidence of the wavefunction collapse. Given that both the definiteness of the measurement outcomes and the quantum-to-classical transition suggest that there should exist some restrictions of what superpositions are physically meaningful, perhaps a promising research direction is to test the limits of superposition and to search for new superselection rules involving interactions \citep{Sto16MicroClassCats}.

\textbf{Acknowledgements.} The author cordially thanks Lawrence S. Schulman and the reviewers for very helpful comments and suggestions.


\end{document}